\documentclass[prd,amsfonts,amsmath,twocolumn,nofootinbib]{revtex4}

\usepackage{nicefrac}

\usepackage{color}
\usepackage{graphicx} 
\usepackage{slashed}
\usepackage{epstopdf}%

\newcommand {\be} {\begin{equation}} 
\newcommand {\ba}{\begin{eqnarray}} 
\newcommand {\ee} {\end{equation}} 
\newcommand{\ea} {\end{eqnarray}}

\renewcommand{\epsilon}{\varepsilon}

\begin{document}

\title{New Physics and the Proton Radius Problem}

\author{Carl E. Carlson}

\author{Benjamin C. Rislow}

\affiliation{Department of Physics, College of William and Mary, Williamsburg, VA 23187, USA}

\date{June 15, 2012}

\begin{abstract}
{\bf Background:}  The recent disagreement between the proton charge radius extracted from Lamb shift measurements of muonic and electronic hydrogen invites speculation that new physics may be to blame.  Several proposals have been made for new particles that account for both the Lamb shift and the muon anomalous moment discrepancies.  {\bf Purpose:}  We explore the possibility that new particles' couplings to the muon can be fine-tuned to account for all experimental constraints.  {\bf Method:}  We consider two fine-tuned models, the first involving new particles with scalar and pseudoscalar couplings, and the second involving new particles with vector and axial couplings.  The couplings are constrained by the Lamb shift and muon magnetic moments measurements while mass constraints are obtained by kaon decay rate data.  {\bf Results:}  For the scalar-pseudoscalar model, masses between 100 to 200 MeV are not allowed.  For the vector model, masses below about 200 MeV are not allowed.  The strength of the couplings for both models approach that of electrodynamics for particle masses of about 2 GeV.  {\bf Conclusions:}  New physics with fine tuned couplings may be entertained as a possible explanation for the Lamb shift discrepancy.

\end{abstract}

\maketitle



\section{Introduction}			\label{sec:intro}


The recent measurement of the muonic hydrogen Lamb shift~\cite{Pohl:2010zza} yielded a proton charge radius $5\sigma$ smaller than the 2006 CODATA value available at the time of its publication~\cite{Mohr:2008fa}, and $7\sigma$ smaller than the 2010 CODATA update~\cite{Mohr:2012tt}, which incorporates the latest proton radius determinations from electron scattering~\cite{Bernauer:2010wm}.  The data used in the CODATA determinations is all electronic.  A possible explanation for the surprising muonic result is that overlooked Standard Model processes or new physics are at present being wrongly attributed to proton size effects.  Assuming the previous electronic measurments of the proton charge radius are correct, alternative explanations must lower the muonic Lamb shift by 310 $\mu$eV to match the experimental result.

Several new physics proposals have been considered to explain the discrepancy.  Jaeckel and Roy~\cite{Jaeckel:2010xx}, as part of a larger investigation into deviations of Coulomb's law, determined that hidden photons that coupled equally to electrons and muons could not explain the discrepancy as they would actually cause the proton radius to appear smaller in ordinary hydrogen.  Tucker-Smith and Yavin~\cite{TuckerSmith:2010ra} developed two simple models in which either a new scalar or vector particle couples to muons and protons.  The strength of the particle couplings were set by constraints placed by the small discrepancy of the muon's anomalous magnetic moment.  They showed that in first order of nonrelativistic perturbation theory, exchange particle masses of order MeV could produce the observed Lamb shift, albeit exchanges with masses this light run afoul of neutron scattering data if the neutron and proton have similar coupling.  Barger \textit{et al.}~\cite{Barger:2010aj} also considered new scalar and vector particles, but suggested it would be difficult for them to satisfy additional constraints placed by $\Upsilon$, $J/\psi$, $\pi$, and $\eta$-decays.

Batell \textit{et al.}~\cite{Batell:2011qq} revived the possibility that a hidden photon could be responsible for the Lamb shift discrepancy by requiring it to couple only to right-handed muons.  This boson also mixes with the photon so the couplings contained additional model dependence not seen in other proposals.   In order to account for the muon anomalous moment constraint, they were forced to introduce and fine-tune the mass of a new scalar particle.  In a second paper Barger \textit{et al.}~\cite{Barger:2011mt} noted that Batell \textit{et al.}'s model does not respect the constraint placed by $K$-decay~\cite{Pang:1989ut} if the decay of their hidden photon were invisible.

In this work we explore the possibility that fine-tuned particle couplings, free from the phenomenological demands of hidden photons, can satisfy muon anomalous moment and $K$-decay, as well as other, constraints.  We consider two separate possibilities.  The first contains two new particles that interact with muons and protons through fine-tuned scalar and pseudoscalar couplings, respectively.  The second contains two new particles that interact with muons and protons through fine-tuned polar and axial vector couplings, respectively.

Our evaluation of the particle mass and coupling parameters proceeds as follows.  We begin in Sec.~\ref{sec:lamb} by finding what coupling parameters are needed to obtain an extra 310 $\mu$eV muonic hydrogen Lamb shift from the exchange of an electrophobic spin-0 or spin-1 particle of a given mass.  In the nonrelativistic limit, pseudoscalar and axial vector particles do not contribute significantly to this shift and their couplings remain free parameters.  Then in Sec.~\ref{sec:mmm} we confront our models to the constraint for the muon anomalous magnetic moment.  Polar vector and axial vector exchange give opposite contributions to the magnetic moment, as do scalar and pseudoscalar exchanges, so cancellations can be arranged if the pseudoscalar and axial couplings are tuned.  We further consider in Sec.~\ref{sec:kdk} the consequences for the decay of $K$'s to two or more unobserved particles.  This decay is possible as a radiative correction to $K\to\mu\nu$ if there is a coupling of a light new particle to the muon, and strong experimental limits are known.  In Sec.~\ref{sec:end} we make some final remarks.


\section{Lamb shift}			\label{sec:lamb}


For a case where there are scalar and pseudoscalar particles coupled to a muon and proton,
\begin{align}
\label{Ls}
{\mathcal L}_S =  &- C^\mu_S \phi \bar{\psi}_\mu  \psi_\mu - iC^\mu_P \varphi \bar{\psi}_\mu \gamma_5 \psi_\mu
\nonumber \\
 	&- C^p_S \phi  \bar{\psi}_p \psi_p	- iC^p_P \varphi  \bar{\psi}_p  \gamma_5 \psi_p,
\end{align}
where $\phi$ is a scalar and $\varphi$ is a pseudoscalar field.
The potential in the nonrelativistic limit is
\begin{equation}
\label{NRs}
V(r) =  - \frac{C_S^\mu C_S^p}{4\pi r} e^{-Mr}		,
\end{equation}
where $M$ is the mass of the exchanged particle, $\phi$.  

The pseudoscalar contributions are much smaller at low momentum transfer.
For the $2P$-$2S$ splitting in hydrogen one gets an energy difference~\cite{Pachucki:1996zza,TuckerSmith:2010ra},
\begin{equation}
\Delta E({\rm 2S-2P}) = - \frac{C_S^\mu C_S^p}{4\pi} 
	\frac{M^2 (m_r\alpha)^3}{2 \left(M+m_r\alpha \right)^4}	\,,
\end{equation}
where (for the muonic case) $m_r = m_\mu m_p/(m_\mu + m_p)$.

\begin{figure}[htpb] 
\begin{center}
\includegraphics[scale=0.65]{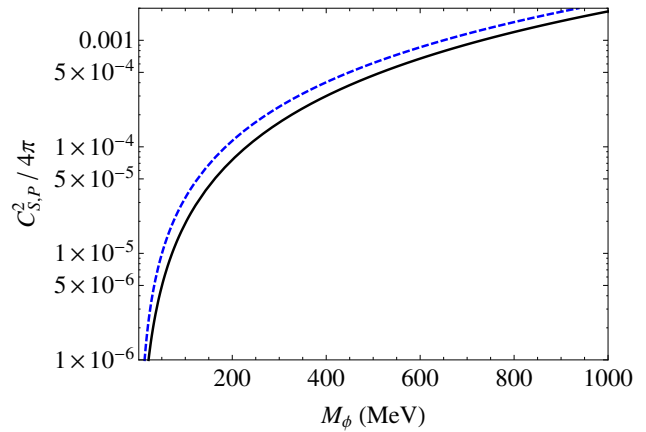}
\caption{The scalar and pseudoscalar couplings needed to satisfy the experimental constraints.  The scalar coupling (solid line) is required to give an extra 310 $\mu$eV to the muonic hydrogen 2S-2P Lamb shift.  The dashed line is the pseudoscalar coupling needed to satisfy the constraint placed by the muon anomalous moment.  We assume the two particle masses are identical.}
\label{fig:Scoupling}
\end{center}
\end{figure}

The scalar coupling required to give an extra 310 $\mu$eV to the muonic hydrogen $2S$-$2P$ Lamb shift, in the case $C_S^\mu = C_S^p =C_S$, is shown by the solid line  in Fig.~\ref{fig:Scoupling} as a function of the exchanged mass, up to an exchanged mass of 1000 MeV.

For the case of polar vector and axial vector particles coupling to the muon and proton,
\begin{align}
\label{Lv}
{\mathcal L}_V =  &- C^\mu_V \phi^\nu \bar{\psi}_\mu \gamma_\nu \psi_\mu - C^\mu_A \varphi^\nu \bar{\psi}_\mu \gamma_\nu \gamma_5 \psi_\mu
\nonumber \\
 	&- C^p_V \phi^\nu  \bar{\psi}_p \gamma_\nu \psi_p	- C^p_A \varphi^\nu  \bar{\psi}_p \gamma_\nu \gamma_5 \psi_p.
\end{align}
where $\phi^\nu$ is a polar voector and $\varphi^\nu$ is an axial vector field.

\begin{figure}[htpb] 
\begin{center}
\includegraphics[scale=0.65]{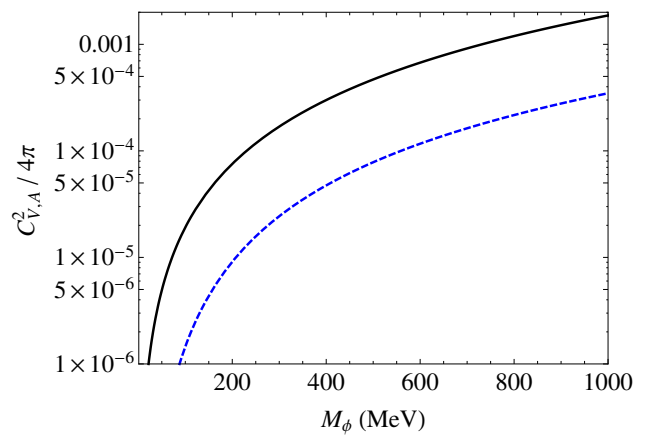}
\caption{The polar and axial vector couplings needed to satisfy the experimental constraints.  The vector coupling (solid line) is required to give an extra 310 $\mu$eV to the muonic hydrogen 2S-2P Lamb shift.  The dashed line is the axial coupling needed to satisfy the constraint placed by the muon anomalous moment.  We assume the two particle masses are identical.}
\label{fig:Vcoupling}
\end{center}
\end{figure}

In the nonrelativistic limit the potential is similar to Eq.~(\ref{NRs}), with vector couplings taking the place of the scalar ones and opposite overall sign.  The solid line in Fig.~\ref{fig:Vcoupling} displays the vector coupling strength as a function of the exchanged mass.

As a side note, these results agree with the known Weak interaction or $Z$-boson exchange contribution to the Lamb shift of a given $nl$ state of hydrogen~\cite{Eides:2000xc},
\begin{equation}
\Delta E^Z_{nl} = - \frac{\alpha (Z\alpha)^3 m_r}{\pi n^3} 
	\frac{8 G_F m_r^2}{\sqrt{2} \alpha}   
	\left( \frac{1}{4}-\sin^2\theta_W \right)^2
	\delta_{l0}		\,,
\end{equation}
with the appropriate substitutions.  (Just the vector part of the $Z$ interaction contributes; further $M\to M_Z$,
\begin{equation}
C_S^p \to -C_S^\mu \to \frac {g}{2 \cos\theta_W} g_V^p	
	= \frac {g}{2 \cos\theta_W} \left(\frac{1}{2} - 2 \sin^2\theta_W \right)   ,
\end{equation}
and $G_F/\sqrt{2} = g^2/(8M_Z^2 \cos^2\theta_W)$.  The numerical result for the $n=2$ levels is about a million times smaller than the extra muonic Lamb shift, speaking to the fact that to get an effect relevant to the Lamb shift problem with an exchange particle mass in the 90 GeV range would require a coupling much stronger that the Weak coupling, even without the $(1/4 - \sin^2\theta_W)$ factors.

The exchange of a Higgs boson also cannot account for the Lamb shift discrepancy.  The energy shift due to a Higgs boson of mass 125 GeV is roughly eight orders of magnitude smaller than what is observed.


\section{Muon magnetic moment}			\label{sec:mmm}


The muon anomalous moment is accurately measured.  The theory for the anomalous moment is also quite accurate, with the bulk of the error coming from uncertainties in hadronic contributions.  There is a small but persistent discrepancy between experiment and theory.  In terms of $a_\mu = (g-2)_\mu/2$,
\begin{align}
a_\mu(\rm data) &= (116\ 592\ 089 \pm 63) \times 10^{-11}
		\quad [0.5 {\rm\ ppm}],
\nonumber\\
a_\mu(\rm thy.) &= (116\ 591\ 840 \pm 59) \times 10^{-11}
		\quad [0.5 {\rm\ ppm}],
\nonumber\\
\delta a_\mu &= (249 \pm 87) \times 10^{-11}
	\quad [2.1 {\rm\ ppm} 	\pm 0.7 {\rm\ ppm}].
\end{align}
The data is from~\cite{Bennett:2004pv,Roberts:2010cj} and the latest theory number is from~\cite{Aoyama:2012wk}.

\begin{figure}[htbp]
\begin{center}
\includegraphics{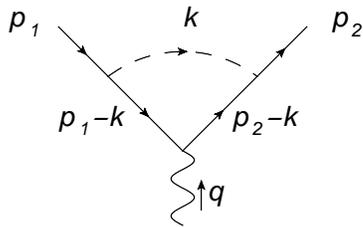}
\caption{One-loop magnetic moment correction}
\label{fig:mmm}
\end{center}
\end{figure}

This discrepancy is four orders of magnitude in fractional terms smaller than the one due to the Lamb shift.  Every particle that contributes to the Lamb shift also contributes to the magnetic moment at the one loop level, as in Fig.~\ref{fig:mmm}.   The contributions of the  pseudoscalar and axial vector, whose couplings are not constrained by the Lamb shift,  have opposite sign to those from the scalar and polar vector, and can be tuned to respect this much smaller discrepancy.  


For scalar and pseudoscalar particles, we consider their masses to be the same.  The magnetic moment result is known in the literature~\cite{Leveille:1977rc,McKeen:2009ny},
\begin{align}
\label{scalmom}
\delta a_\mu 
	&= 
	\frac{m_\mu^2}{8\pi^2}\int_0^1		dz   \ 
	\frac{C_S^2 \,	z^2(2-z)	-C_P^2 \,  z^3} {z^2m_\mu^2 +(1-z)M^2}
	\nonumber\\
&
=\frac{1}{8\pi^2}\big[C_S^2 H_S(r) - C_P^2 H_P(r)\big],
\end{align}
where $r = M^2/m_\mu^2$,
\begin{align}
H_S(r) &=   \frac{3-2r}{2} + \frac{r(r-3)}{2} \ln r
	\nonumber\\
	&- (r-1) \sqrt{r(r-4)}\ \ln\left[ \frac{\sqrt{r}+\sqrt{r-4}}{2} \,\right] 
\end{align}
and
\begin{align}
H_P(r) &=  - \frac{2r+1}{2} + \frac{r(r-1)}{2} \ln r
	\nonumber\\
	&- \frac{r^{3/2} (r-3)}{\sqrt{r-4}} \ln\left[\frac{\sqrt{r}+\sqrt{r-4}}{2} \,\right].
\end{align}
(The expressions continue nicely to $r<4$.)  Low and high mass limits are
\begin{align}
\delta H_S = \left\{
	\begin{array}{ll}
	\frac{ 3  }{ 2  } \ ,	& M \ll m_\mu	\,, 	\\[1.25ex]
	\ln\frac{M^2}{m_\mu^2} 
		- \frac{7}{6}  ,
	& M \gg m_\mu		\,,
	\end{array}
	\right.
\end{align}
and
\begin{align}
\delta H_P = \left\{
	\begin{array}{ll}
	 \frac{  1  }{ 2  } \ ,	& M \ll m_\mu	\,, 	\\[1.25ex]
	 \ln\frac{M^2}{m_\mu^2} - \frac{11}{6}  ,
	& M \gg m_\mu		\,.
	\end{array}
	\right.
\end{align}
Eq.~(\ref{scalmom}) can be rearranged to solve for $(C^\mu_P)^2$.  The result is plotted as the dashed line in Fig.~\ref{fig:Scoupling}.    One notices that fine tuning must be done to several significant figures at higher masses.

For polar and axial couplings, we also only consider the case where their masses are equal.  Their contribution to the muon's magnetic moment is
\begin{align}
\label{vecmom}
\delta a_\mu &=\frac{m_\mu^2}{4\pi^2}\int_0^1 \frac{dz}{z^2m_\mu^2 +(1-z)M^2}\bigg\{C_V^2 z^2(1-z)
\nonumber \\
&	\hskip 17 mm
-C_A^2 \bigg[z(1-z)(4-z)+\frac{2m_\mu^2}{M^2}z^3\bigg]\bigg\}
	\nonumber\\
&= \frac{1}{4\pi^2} \left[ C_V^2 H_V(r) - C_A^2 H_A(r) \right]	\,.
\end{align}
Here~\cite{Leveille:1977rc,McKeen:2009ny}
\begin{align}
H_V(r) &= \frac{1-2r}{2} + \frac{r(r-2)}{2} \ln r
			\nonumber\\
	&- \frac{r^{1/2} (r^2-4r+2)}{\sqrt{r-4}} \ln\left(\frac{\sqrt{r}+\sqrt{r-4}}{2}\right)  ,
\end{align}
with limits
\begin{align}
H_V(r) = \left\{
	\begin{array}{cl}
	\frac{m_\mu^2}{3M^2}	\,, 	& M \to \infty	\,,	\\[1.0ex]
	\frac{1}{2}			\,,	& M \to 0		\,,
	\end{array}
	\right.
\end{align}
and~\cite{Leveille:1977rc}
\begin{align}
H_A(r) &= \frac{1}{r} + \frac{2r-5}{2} - \frac{r^2 - 4r+2}{2} \ln r
			\nonumber\\
	&+ (r-2)\sqrt{r(r-4)} \ln\left(\frac{\sqrt{r}+\sqrt{r-4}}{2}\right)  ,
\end{align}
with
\begin{align}
H_A(r) = \left\{
	\begin{array}{cl}
	\frac{5m_\mu^2}{3M^2}	\,, 	& M \to \infty	\,,	\\[1.0ex]
	\frac{m_\mu^2}{M^2}	-\ln \frac{M^2}{m_\mu^2} 
		- \frac{5}{2} + \ldots		\,,	& M \to 0		\,.
	\end{array}
	\right.
\end{align}

Rearranging  Eq.~(\ref{vecmom}) allows for the evaluation of $(C_A^\mu)^2$.  The result is plotted as the dashed line in Fig.~\ref{fig:Vcoupling}.

More general combinations of $S$, $P$, $V$, and $A$ are also possible.  Of note is the model of Ref.~\cite{Batell:2011qq}, which involves a vector particle with extra parity violating coupling to the muon with (in our notation) fixed $C_V^\mu$ and $C_A^\mu$, that achieves fine tuning using a scalar, also with definite muonic coupling, but with a tunable mass.


\section{$K$ decay with unobserved neutrals}			\label{sec:kdk}

\begin{figure}[htbp]
\begin{center}
\includegraphics{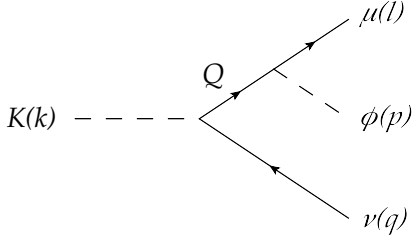}
\caption{Kaon decay with an extra neutral scalar, $\phi$, either scalar or pseudoscalar.}
\label{fig:kdk}
\end{center}
\end{figure}

If a light neutral particle couples to muons, the decay $K\to\mu\nu\phi$ is possible, Fig.~\ref{fig:kdk}.  There has been an experimental search for multibody decays $K\to\mu X$ where $X$ contains only neutral particles that are not photons~\cite{Pang:1989ut}.  The experiment searched for muons in the kinetic energy range 60 to 100 MeV---for reference, the muon from $K\to\mu\nu$ has $T_\mu = 152$ MeV or $E_\mu = 258$ MeV---and found a strong limit.  To give the result with some precision, the experimenters give their detector efficiency function $D(E_\mu)$, which is the relative efficiency to detect an energy $E_\mu$ muon compared to a 258 MeV muon, and which is zero outside the stated limits and smoothly varying in between.


For a given decay spectrum $d\Gamma(K\to\mu X)/dE_\mu$, the experimental limit is quoted as~\cite{Pang:1989ut}
\begin{align}
	\label{eq:limit}
\frac{1}{\Gamma(K\to\mu\nu)} \int \frac{d\Gamma(K\to\mu X)}{dE_\mu}
	D(E_\mu) \, dE_\mu  <  2 \times 10^{-6}  \,.
\end{align}
The simple $K_{\mu2}$ decay rate is
\begin{align}
\Gamma(K\to \mu\nu) =  \frac{ G_F^2 f_K^2 V_{us}^2}{4\pi m_K^3}
	m_\mu^2 \big(m_K^2 - m_\mu^2 \big)^2		\,.
\end{align}
where the kaon decay constant is defined from
\begin{equation}
\langle 0 | \ \bar u \, \gamma_\mu (1-\gamma_5) s \ | 0 \rangle 
	= \sqrt{2} f_K k_\mu \,.
\end{equation}

Generically, a three body decay is given by
\begin{align}
\Gamma(K\to\mu\nu\phi) = \frac{1}{64\pi^3 m_K} \int dE_\mu \, dE_\nu
	\sum_{\rm spins} | \mathcal M |^2		\,,
\end{align}
with integration limits
\begin{equation}
m_\mu \le E_\mu \le \frac{m_K^2 + m_\mu^2 -m_\phi^2}{2m_K}	\,,
\end{equation}
and
\begin{align}
\Big\{		\begin{array}{r}
		max \\
		min
		\end{array}		\Big\} E_\nu
= \frac{m_K^2 + m_\mu^2 - m_\phi^2 -2m_K E_\mu}
	{2 \Big( m_K-E_\mu \mp \sqrt{E_\mu^2-m_\mu^2} \,\Big)}		\,.
\end{align}

The matrix element for the decay into a muon, neutrino, and particle with both scalar and pseudoscalar couplings is
\begin{align}
&\mathcal M_{S,P} 
= 	\frac{G_F f_K V_{us}}{Q^2-m_\mu^2}  \bar u(l) \nonumber\\
&\ \times	\left[ (C^\mu_S - i C^\mu_P) Q^2 
	+ m_\mu (C^\mu_S + i C^\mu_P) \!\not\! k \right]
	(1-\gamma_5) v(q)		\,,
\end{align}
where $Q^2 = (k-q)^2 = m_K^2-2m_K E_\nu$.

The matrix element squared and summed is
\begin{align}
\sum_{\rm spins} | \mathcal M_{S,P} |^2 
	&= \frac{4 G_F^2 f_K^2 V_{us}^2}{(Q^2-m_\mu^2)^2}
	\bigg\{ ({C^\mu_S}^2 + {C^\mu_P}^2) 	\nonumber\\
&\times	\Big[ 2m_K E_\mu Q^2 (Q^2-m_\mu^2)  \nonumber\\
&	- (Q^4 - m_\mu^2 m_K^2)(Q^2+m_\mu^2 -m_\phi^2)
	\Big]		\nonumber\\
	& +
	2 ({C^\mu_S}^2 - {C^\mu_P}^2) \, m_\mu^2 Q^2 (m_K^2 - Q^2)  \bigg\}	\,.
\end{align}
We evaluate the left-hand-side of Eq.~(\ref{eq:limit}) using our constrained 
$C^\mu_S$ and $C^\mu_P$ couplings for a given scalar or pseudoscalar mass 
$m_\phi$.  We note that neither the anomalous moment nor the square of the matrix element contain terms with both $C^\mu_S$ and $C^\mu_P$.  Thus, a model with two equal mass scalar and pseudoscalar particles is indistinguishable from one which has only a single particle with both scalar and pseudoscalar couplings. 

\begin{figure}[tbp]
\begin{center}
\includegraphics[scale=.65]{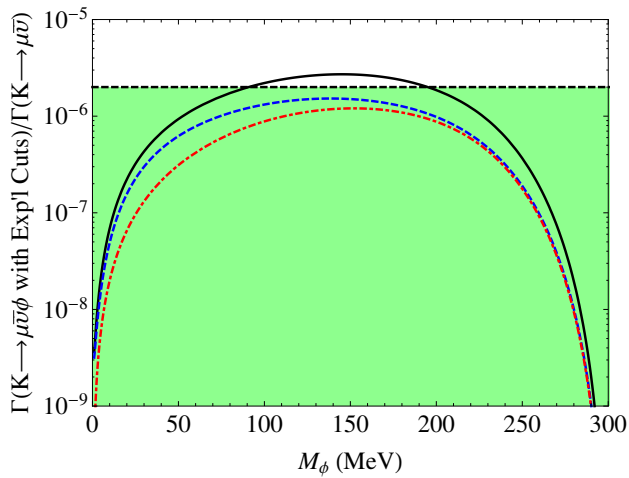}
\caption{Mass limits on scalar and pseudoscalar particles due to constraints placed by $K\to\mu X$ searches.  The solid curve is the full result, accounting for the experimental efficiency, obtained through satisfying the Lamb shift and magnetic moment criteria.  The contributions of the scalar (dashed curve) and pseudoscalar (dash-dotted curve) couplings are indicated separately.  The experimental limit is the horizontal line, and the shaded region is allowed.  }
\label{fig:Klimit}
\end{center}
\end{figure}

Comparison of our calculated experimental-efficiency-weighted decay rate to the experimental limit is shown in Fig.~\ref{fig:Klimit}.  A range of scalar masses from about 100 to 200 MeV is not allowed.  For other masses, the couplings are not excluded.


The matrix element for the decay into a muon, neutrino, and a particle with both polar and axial vector couplings is, using $C_{R,L}^\mu = C_V^\mu \pm C_A^\mu$
\begin{align}
\mathcal M_{V,A} 
&= 	\frac{G_F f_K V_{us}}{Q^2-m_\mu^2} \epsilon_\nu \bar u(l) \gamma^\nu \nonumber\\
&	\times	\left[ C_L^\mu Q^2 
	+ m_\mu C_R^\mu \!\not\! k \right]
		(1-\gamma_5) v(q)		\,,
\end{align}
where $\epsilon_\nu$ is the polarization vector of the new particle.  This leads to
\begin{align}
&\frac{d\Gamma(K\to \mu\nu V)}{dE_\mu \, dE_\nu} =  
	\frac{ \Gamma(K\to \mu\nu) }{4\pi^2 m_\mu^2 \big(m_K^2 - m_\mu^2 \big)^2}
	\frac{m_K^2} {\big( Q^2-m_\mu^2 \big)^2}		\nonumber\\
&	\times	
	\bigg\{ 4 {C^\mu_R}^2 m_\mu^2 m_K^2 E_\mu E_\nu
	- 12 C^\mu_R C^\mu_L m_\mu^2 m_K Q^2 E_\nu	\nonumber\\
&	+ \left[ {C^\mu_L}^2 Q^4 - {C^\mu_R}^2 m_\mu^2 m_K^2 \right]
		\left(m_K^2 + m_V^2 - m_\mu^2 - 2m_K E_V \right)	\nonumber\\
&
	+ \frac{1}{m_V^2} \left(m_K^2 - m_V^2 - m_\mu^2 -2m_K E_v \right)
\Big[4 {C^\mu_R}^2 m_\mu^2 m_K^2 E_V E_\nu
				\nonumber\\
	& 	+ \left( {C^\mu_L}^2 Q^4 \! -  {C^\mu_R}^2 m_\mu^2 m_K^2 \right)
	\left(m_K^2 \! - \! m_V^2 \! + \! m_\mu^2 \! - \! 2m_K \! E_\mu \right)	 \Big]
	\bigg\}		.
\end{align}

We note that this matrix element squared contains terms with $C^\mu_V C^\mu_A$. 
Hence one must distinguish the case of two particles of equal mass from one parity violating particle with polar and axial couplings.  The anomalous moment has no cross terms, so both the one and two particle cases contain the same couplings. However, in the decay process, the parity violating case has the possibility of destructive interference.


\begin{figure}[tbp]
\begin{center}
\includegraphics[scale=.65]{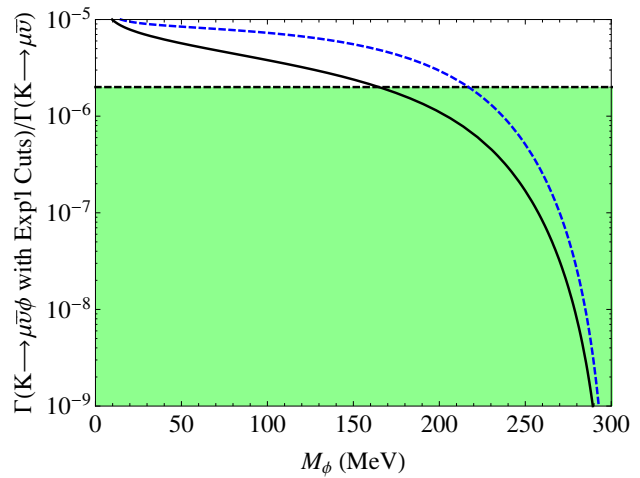}
\caption{Mass limits on polar and axial vector particles due to constraints placed by $K\to\mu X$ searches.  The solid curve is the result for a single particle with both polar and axial vector couplings, accounting for the experimental efficiency, with couplings obtained through satisfying the Lamb shift and magnetic moment criteria.  The dashed curve is the result for separate polar and axial vector particles with equal masses.  The experimental limit is the horizontal line, and the shaded region is allowed.  }
\label{fig:Klimit2}
\end{center}
\end{figure}


The results for both cases are shown in Fig.~\ref{fig:Klimit2}, using couplings obtained from the fine tuning of the muon anomalous moment.  For the one particle case, we show results for $C_V^\mu$ and $C_A^\mu$ having the same sign.  Masses below about 160 MeV are not allowed for this scenario.  For two particles with equal masses, the disallowed region extends to about 210 MeV.  All higher masses are allowed.


\section{Closing Comments}			\label{sec:end}

Exotic, in the sense of presently undiscovered, particles that couple to muons and hadrons but not electrons could be responsible for an extra energy shift in muonic hydrogen, and thereby reconcile the muonic and electronic proton radius measurements.  However, for exotic explanations to work, there are requirements to be met, a number of which have been discussed already.

So far we have not confronted all the additional constraints placed by neutron scattering and meson decays mentioned in~\cite{Barger:2010aj}.  To note some of these briefly, there are decay constraints following from searches for unknown particles in $\Upsilon$ or $J/\psi$ decays.  These are potentially serious.  However, they do not apply if there is no coupling to heavier quarks.  Hence a new force or new particle that spoofs a smaller proton radius in muonic hydrogen should couple only or almost only to second generation leptons and first generation hadrons.  In particular, there can be coupling to muons but not electrons and to first generation quarks but not $b$ or $c$ quarks.

Further noted in~\cite{Barger:2010aj}, neutron scattering constraints only limit models with very light new particle masses (under $5$ MeV), other muonic atom energy splittings give bounds already below ones discussed, the $\pi \to \gamma V$ decay where $V$ is a massive vector only impacts an $m_V$ mass range that is already excluded, and the limits from non-observance of the decay $\eta \to V V$ is not serious for $m_V$ below $m_\eta / 2$ given the results in Figs.~\ref{fig:Scoupling} and~\ref{fig:Vcoupling}.

Fine tuning is needed to evade constraints from the muon magnetic moment measurements.  Tuning is possible because polar vector and axial vector exchanges give opposite sign contributions to $(g-2)_\mu$, and the same is true of scalar and pseudoscalar exchanges.   The need for fine tuning has the additional effect that once a vector (or scalar) exchange contributing to the Lamb shift requires a further axial (or pseudoscalar) exchange in $(g-2)_\mu$, the axial vector (or pseudoscalar) will then potentially contribute to other processes, adding for example to the rate that needs to be evaded in a decay process. 

In models where the new physics proton and muon couplings have similar magnitudes, $K$-decay constraints must be avoided by having the new particle mass above at least $150$ MeV for the polar vector-axial vector equal mass case, or by having mass low, below $100$ MeV or high, above $180$ MeV, in the scalar-pseudoscalar case.

Relatively new, much from the past year, are limits arising from searches for dark photons, which are massive vector particles that couple to charged particles at a reduced rate.  The combination of KLOE~\cite{Giovannella:2011nh}, APEX~\cite{Abrahamyan:2011gv}, Mainz A1~\cite{Merkel:2011ze}, and BABAR~\cite{Aubert:2009cp} limit a new vector coupling to below about a few $\times\ 10^{-3}$ of the normal photon coupling for the $m_V$ mass range $60$ to above $500$ MeV, and below $10^{-3}$ over part of that range.   This is relevant to eliminating proposed theories where the magnitudes of the electron and proton couplings are the same or similar.  Then to give a muonic hydrogen energy sufficient to spoof the proton radius result would require a muon coupling so large as to violate the $K$-decay bound by more than an order of magnitude, at least in the mass range $60$ to $310$ MeV (where the phase space for the relevant experiment runs out).

Our present work should be viewed as a proof of concept rather than a completed model.  For successful exotic explanations of the proton radius problem, there are requirements of coupling only to targeted leptons and hadrons, of fine tuning the muon couplings, and of restricted mass ranges to avoid conflict with unobserved decays.  While these requirements may seem difficult, there is still a window of possibility for new physics explanations of the proton radius problem.  One would like an \textit{ab initio} new physics theory that works for this problem, and the restrictions given here may be a help in finding one.


\begin{acknowledgments}

We thank the National Science Foundation for support under Grant PHY-0855618 and thank Will Detmold an Gunnar Ingelman for helpful conversations.

\end{acknowledgments}

\bibliography{NewPhysProRadius}

\end{document}